\documentstyle[prl, aps, twocolumn, epsf]{revtex}

\begin{document}

\title{Susceptibility and Low Temperature Thermodynamics of
 Spin-1/2 Heisenberg Ladders}
\author{Beat Frischmuth$^{a,b}$, Beat Ammon$^{a,b}$ and
 Matthias Troyer$^{a,b,c}$}
\address{$^{a}$Theoretische Physik, Eidgen\"ossische Technische
Hochschule, CH-8093 Z\"urich, Switzerland\\
$^{b}$Swiss Center for Scientific Computing, Eidgen\"ossische
 Technische Hochschule, CH-8092 Z\"urich, Switzerland\\
$^{c}$Institute for Solid State Physics, Univerisity of Tokyo,
 Roppongi 7-22-1, Minatoku, Tokyo 106, Japan}

\maketitle
\begin{abstract}
The temperature dependence of the uniform susceptibility and the
ground state energy of antiferromagnetic Heisenberg ladders with up to
6 legs has been calculated, using the Monte Carlo loop algorithm. The
susceptibilities of even-leg-ladders show spin gaps while these of
odd-leg-ladders remain finite in the zero temperature limit. For
small ratios of intra- to inter-leg couplings, odd-leg-ladders can be
mapped at low temperatures to single chains. For equal couplings, the
logarithmic corrections at low temperatures increase markedly with the
number of legs.
\end{abstract}

\vspace{0.5cm}  
Recently, antiferromagnetic Heisenberg spin-$1/2$ ladder systems have
attracted much interest, following the discovery of a finite spin gap
in the 2-leg-ladder \cite{dagottoriera}. Later investigations showed
that the crossover from the single Heisenberg chain to the
two-dimensional (2D) antiferromagnetic square lattice, obtained by
assembling chains to form ``ladders'' of increasing width, is far from
smooth \cite{dagotto}. Heisenberg ladders with an even number of legs
(chains), $n_{l}$, show a completely different behavior than
odd-leg-ladders. While even-leg-ladders have a spin gap and short
range correlations, odd-leg-ladders have no gap and power-law
correlations. Based on density matrix renormalization group (DMRG)
studies, White et al. \cite{white} gave an explanation of this
fundamental difference in the framework of the Resonant Valence Bond
(RVB) picture \cite{anderson}. These theoretical predictions have been
verified experimentally, in materials such as $(VO)_{2}P_{2}O_{7}$
\cite{johnston} and the homologous series of cuprates
$Sr_{n-1}Cu_{n+1}O_{2n}$ \cite{hiroi}, which contain weakly coupled
arrays of ladders.

Previous numerical studies, using exact diagonalization
\cite{dagottoriera,poilblanc}, the Quantum Monte Carlo world line
algorithm \cite{barnes} or the Quantum Transfer Matrix method
\cite{troyer} were restricted to small systems or could not be applied
at low temperatures. Using the new Quantum Monte Carlo (QMC) loop
algorithm \cite{evertz}, we overcome these limitations and are able to
investigate very large ladders to much lower temperatures.

In this letter we consider ladders with $n_{l}$ legs
($n_{l}=1,2,\ldots,6$) of length $L$. The Hamiltonian of such
spin$-1/2$ systems

\begin{equation}
H=J \sum_{\leftrightarrow} \vec{S_{i}} \vec{S_{j}}
    + J_{\perp}\sum_{\updownarrow} \vec{S_{i}} \vec{S_{j}}
\end{equation}
is defined on $L\:\times\:n_{l}$ lattices. The sum marked by
$\leftrightarrow$ ($\updownarrow$) runs over nearest neighbors along
the legs (rungs). We assumed periodic boundary conditions in the
longitudinal direction of the ladder. The exchange constants $J$ and
$J_{\perp}$ are positive, corresponding to antiferromagnetic coupling.

Using the Quantum Monte Carlo loop algorithm with improved estimators,
we have calculated the temperature dependence of the uniform
susceptibility $\chi$ and the internal energy $E$. The QMC loop
algorithm was first developed by Evertz et al. \cite{evertz} for the
$6$-vertex-model, but can also be applied to quantum spin systems
\cite{evertz2,wiese}. The QMC loop algorithm is an improved world line
algorithm. The updates in the loop algorithm are global and no longer
local as in the conventional Metropolis world line algorithm. This has
the great advantage that the autocorrelation times are reduced by
several orders of magnitude. It was thus possible to simulate very
long ladders to very low temperatures. We considered systems up to 100
x 6 sites and reached temperatures down to $T=J/50$ without major
problems. All results are extrapolated to a Trotter time interval
$\Delta\tau\rightarrow 0$. The application of improved estimators (see
e.g. Ref. \cite{wiese}), further reduces the variance of the measured
observables dramatically.

In Table \ref{ground} the ground state energies of the different
ladders in the isotropic case ($J=J_{\perp}$) are
presented. Considering ladders of lengths $L$, we first extrapolate to
$T\rightarrow 0$. For the odd-leg-ladders we use the form
\begin{equation}\label{odd}
E_{L}(T)=E_{L}(0)+a\:T^{2}, 
\end{equation}
where $E_{L}(T)$ is the internal energy of the ladder of length
$L$. This form is motivated by the infinite single chain, which in a
low temperature field theory can be described by a massless boson. Our
numerical data agree well with Eq. (\ref{odd}) for the finite single
chains, as well as for the finite 3- and 5-leg-ladders. The
even-leg-ladders, on the other hand, have spin gaps. The internal
energy for $T$ well below the spin gap $\Delta$ is determined by the
thermal occupation of the lowest lying $S=1$ magnon band with a
quadratic dispersion near the zone boundary minimum. For the
extrapolation to $T\rightarrow 0$ we therfore use the form
$E_{L}(T)=E_{L}(0)+b\:(T^{3/2}+2\Delta T^{1/2})\exp(-\Delta/T)$. In a
second step the ground state energies $E_{L}(0)$ for the finite
systems are extrapolated to the bulk limit $E_{L=\infty}(0)$, fitting
$E_{L}(0)$ to a polynomial in $1/L$. The finite size corrections of
the ground state energy, however, are negligibly small for ladders
with $L\geq100$.

In Fig. \ref{Fig1} we show the susceptibility per rung of $n_{l}$
spins, $\chi(T)$, for the ladders with $1\leq n_{l} \leq 6$ in the
isotropic case $J=J_{\perp}$. We always consider ladders long enough
such that finite size corrections are negligible. For $T>J$ the
results agree well with a third order high temperature expansion. At
low temperatures, we observe the predicted behavior. Ladders with even
$n_{l}$ show an exponential drop of the susceptibility indicating a
gap in the excitation spectrum. For larger $n_{l}$, the drop sets in
at smaller $T$ and is steeper. The gap $\Delta_{n_{l}}$ decreases
substantially with increasing $n_{l}$. For odd $n_{l}$ on the other
hand, $\chi(T)$ remains finite also at $T \ll J$, as in the single
chain.

Ladders with even $n_{l}$ were already investigated in detail
\cite{white,troyer,gopalan}. For temperatures $T\ll\Delta$ the
susceptibility for the 2-leg-ladder is determined by the thermal
occupation of $S=1$ magnon band with a quadratic dispersion near the
zone boundary minimum \cite{troyer}:
\begin{equation}
\chi(T) \propto T^{-1/2} e^{-\Delta/T}\label{fit}, \qquad T \ll \Delta.
\end{equation}
Provided a quadratic dispersion for the lowest lying magnon branch in
the excitation spectrum near its minimum is assumed, Eq. (\ref{fit})
also holds for the 4- and 6-leg-ladder. We estimate $\Delta_{n_{l}}$
by fitting the numerical QMC data for low temperatures and find in the
isotropic case $\Delta_{2}=0.51(1)J$ for $n_{l}=2$,
$\Delta_{4}=0.17(1)J$ for $n_{l}=4$ and $\Delta_{6}=0.05(1)J$ for
$n_{l}=6$. The value $\Delta_{2}$ obtained for the 2-leg-ladder is in
perfect agreement with former results \cite{white,barnes,troyer}. On
the other hand, the spin gap obtained by White et al. \cite{white},
using DMRG methods for the 4-leg-ladder $\Delta_{4}=0.190J$ is
slightly larger than our value.

The decrease of the spin gap with increasing $n_{l}$ can be explained
by delocalization of RVB singlets not only along but more and more
also across the ladder. The decrease of the spin gap, however, is much
faster than $\Delta_{n_{l}}\propto 1/n_{l}$, suggested in
\cite{white}. The spin gap for the 6-leg-ladder $\Delta_{6}$ is
already a factor $10$ smaller than $\Delta_{2}$, suggesting rather an
exponential decrease of $\Delta_{n_{l}}$ with increasing $n_{l}$.

The susceptibility per rung of the odd-leg-ladders remains finite in
the low temperature limit and tends to a zero temperature value
approximately independent of $n_{l}$ [see Fig. \ref{Fig1}(b)]. For
this we conclude that the odd-leg-ladders belong to the same
universality class as the single chain.

The single chain can be described in a low temperature field theory by
the $k=1$ Wess-Zumino-Witten nonlinear
$\sigma$-model or equivalently by a free, massless boson, with a
(spin) velocity $v=\pi J/2$. Based on this model $\chi(T=0)$ and the
leading $T$ dependences of $\chi$ have been calculated
\cite{eggert,nomura} with the leading marginally irrelevant
operator. Two-loop renormalization of the marginal coupling leads to
\begin{eqnarray}
\label{1susc}
\chi(T)=&\frac{1}{2\pi v} + \frac{1}{4 \pi v}\left[
\frac{1}{\ln(T_{0}/T)}
-\frac{\ln(\ln(T_{0}/T)+1/2)}
{2\ln^{2}(T_{0}/T)}\right] \nonumber \\
	&	+ O\left((\ln\: T)^{-3}\right),
\end{eqnarray}
where $T_{0}$ is the cut-off-temperature. The susceptibility
approaches its asymptotic zero temperature value \mbox{$\chi(0)=(2\pi
v)^{-1}=(J \pi^{2})^{-1}$} with infinite slope. The field theoretical
results can be compared to the exact Bethe ansatz data \cite{eggert}
and one finds that Eq. (\ref{1susc}) holds to within 1\% for
$T<0.1J$. These results are shown in Fig. \ref{Fig2} together with
our QMC data for low temperatures.

We turn now to the 3- and 5-leg-ladders. In the limit $J/J_{\perp}=0$
each eigenfunction is a direct product of one-rung states whose lowest
lying multiplet is a spin doublet ($s=1/2$). The ground state of the
whole system is therefore $2^{L}$-fold degenerate. A finite value of
$J$ lifts this degeneracy. In this $2^{L}$-dimensional subspace $\cal
M$ we can define an effective Hamiltonian $H_{\mbox{\scriptsize eff}}$
which includes all intra-leg interactions. To first order in
$J/J_{\perp}$ we obtain \cite{hatano}:
\begin{equation}\label{map}
H^{(1)}_{\mbox{\scriptsize eff}}=J_{\mbox{\scriptsize eff}}\:
\sum_{j=0}^{\infty} \vec{S}_{j,\:\mbox{\scriptsize
tot}}\vec{S}_{j+1,\:\mbox{\scriptsize tot}},
\end{equation}
where $\vec{S}_{j,\:tot}$ is the total spin of the $j^{th}$ rung and
$J_{\mbox{\scriptsize eff}}=J$ for the 3-leg-ladder, respectively
$J_{\mbox{\scriptsize eff}}=1.017\:J$ for the
5-leg-ladder. $H^{(1)}_{\mbox{\scriptsize eff}}$ has just the form of
the Hamiltonian of the single chain with an effective coupling
$J_{\mbox{\scriptsize eff}}$ and we can map the low lying energy
states of the 3- and 5-leg-ladder to those of the single chain. In the
following we will concentrate on the 3-leg-ladder.

The susceptibility of the single chain $\chi_{1}(T/J)$ scales with
$1/J$. It follows, that for a 3-leg-ladder with small $J/J_{\perp}$
and at low temperature, where only the above mentioned low lying
states in $\cal M$ are relevant, the susceptibility per rung
$\chi_{3}$ scales with $1/J_{\mbox{\scriptsize eff}}$ and has the same
functional dependence on $T/J_{\mbox{\scriptsize eff}}$ as
$\chi_{1}(T/J)$, according to Eq. (\ref{map}). This can be seen in
Fig. \ref{Fig3}, where we show $J\chi_{3}$ for different ratios
$J/J_{\perp}$ as a function of $T/J$. For small $J/J_{\perp}$ the
susceptibility per rung $\chi_{3}$ multiplied by $J$ is very close to
$J\chi_{1}$ until a crossover temperature, which depends on
$J/J_{\perp}$. Above this temperature, the susceptibility of the
3-leg-ladder is larger, due to the presence of additional states in
the 3-leg-ladder which are not included in the $2^{L}$-dimensional
subspace $\cal M$. These additional states have a finite gap $\tilde
\Delta$. The susceptibility of the 3-leg-ladder then reads
\begin{equation}
\chi_{3}=\chi_{1}(J_{\mbox{\scriptsize eff}})+\tilde \chi\label{model12},
\end{equation}
where $\tilde \chi$ is the contribution of the additional states. From
our QMC data we find $J_{\mbox{\scriptsize eff}}\approx J$ for all
small $J/J_{\perp}$ and $\Delta_{4}(J/J_{\perp})\lesssim\tilde\Delta
\lesssim\Delta_{2}(J/J_{\perp})$.

 In the isotropic case $J=J_{\perp}$, the gap $\tilde\Delta$ in units
of $J$ is indeed smaller than for small $J/J_{\perp}$ but remains
finite ($\Delta_{4}\lesssim\tilde\Delta\lesssim\Delta_{2}$). For
$T\ll\tilde\Delta$ we can neglect the contribution $\tilde\chi$ of the
additional states. Comparing $\chi_{3}$ to $\chi_{1}$ for
$T\ll\Delta$, we see that their slopes are completely different (see
Fig. \ref{Fig3}). Therefore we conclude that in this case the
simple model, discussed above [Eq. (\ref{model12})], no longer
applies.

As $J/J_{\perp}\rightarrow 1$ also next-nearest neighbor and longer
range interactions between rung-spins become important in the
effective Hamiltonian $H_{\mbox{\scriptsize eff}}$. Since these
additional interactions respect the $SU(2)$ and translational
symmetry, Eq. (\ref{1susc}) still applies for some values of $v$ and
$T_{0}$, according to \cite{eggert}. Therefore, for very low
temperatures the susceptibility of the 3-leg-ladder can be described
by Eq. (\ref{1susc}) also in the isotropic case (see
Fig. \ref{Fig2}).

The spin velocity in the isotropic 3-leg-ladder $v_{3}$ seems to be
close to that of the single chain $v_{1}$. This can be seen by two
facts. First, both of the susceptibilities per rung seem to
extrapolate to the same zero-temperature value. Secondly, White et
al. \cite{white} determined the spin gap of the finite single chain
and the finite 3-leg-ladder in function of $1/L$. The slopes of these
curves as $1/L\rightarrow 0$, $\pi v_{1}$ and $\pi v_{3}$, agree, at
least within 5\%. Assuming $v_{1}=v_{3}$, we get a rough estimate of
the cut-off-temperature $T_{0}$ in the isotropic 3-leg-ladder. The
value is much smaller than in the single chain (see
Fig. \ref{Fig2}). We conclude therefore, that the effective
interactions between the spinons in the 3-leg-ladder are much stronger
than those in the single chain.

We conclude that the odd-leg-ladders belong to the same universality
class as the single chain and can be described in the zero temperature
limit by a k=1 Wess-Zumino-Witten non linear $\sigma$-model with a
spin velocity $v_{n_{l}}$. These velocities seem to have more or less
the same value for all $n_{l}$. With increasing $n_{l}$, however, we
move further away from the conformal point and the logarithmic
corrections, due to the leading marginally irrelevant operator,
increase markedly.

Finally, we want to point out, that the zero temperature value of the
susceptibility per site $\chi^{\mbox{\scriptsize (site)}}_{n_{l}}(0)$
for the odd-leg-ladders decreases with increasing $n_{l}$. Since the
odd-leg-ladders seem to have more or less the same zero temperature
value of the susceptibility per rung, it follows that
$\chi^{\mbox{\scriptsize (site)}}_{n_{l}}(0)\propto 1/n_{l}$. For
$n_{l}\rightarrow\infty$ the zero temperature value
$\chi^{\mbox{\scriptsize (site)}}_{n_{l}}(0)$ therefore goes to zero
for odd $n_{l}$ as well as for even $n_{l}$. The susceptibility per
site of a 2D square lattice, however, is finite for $T=0$. This is
therefore a further example that the crossover from the single chain
to the $2D-$lattice is not a smooth one.

We wish to thank G. Sierra, H.G. Evertz, D. W\"urtz and especially
T.M. Rice for very instructive and stimulating discussions. The
calculations were performed on the Intel Paragon of the ETH
Z\"urich. The work was partially supported by the Schweizerischen
Nationalfond (B.F.) and by an ETH internal grant No. 9452/41-2511.5
(B.A.).

\newpage
\begin{table}
\begin{center}
\begin{tabular}{c c c}
number of legs & obtained ground & reference value\\
 & state energy & \\ 
\hline

1&-0.4432(1)&-0.44315\ldots \cite{bethe}\\
2&-0.5780(2)&-0.578 \cite{barnes}\\
3&-0.6006(3)&-\\
4&-0.6187(3)&-\\
5&-0.6278(4)&-\\
6&-0.635(1)&-\\
2D lattice& &-0.6693(1) \cite{wiese}\\
\end{tabular}
\end{center} \caption[*]{
Ground state energies per site for the different ladders in the
isotropic case. For the single chain we have perfect agreement with
the analytical result $\frac{1}{4}- \ln\:2$ from the Bethe Ansatz
\cite{bethe}. Furthermore, the result for the 2-leg-ladder coincides
with the ground state energy calculated by Barnes et
al. \cite{barnes}, using Lanczos techniques. With increasing width the
results approach to the ground state energy per site of the infinite
2D square lattice, which was calculated by various methods. For an
overview see \cite{manousakis}. The reference value given here was
recently obtained by U.J. Wiese and H.P. Ying \cite{wiese} using the
QMC loop algorithm. \label{ground}}
 
\end{table}

\vspace{2cm} 
\begin{figure}
\caption[*]{Susceptibility as a function of the temperature for the
single chain and the Heisenberg ladders with up to 6 legs. At high
temperatures the result agree well with a third order high temperature
expansion. The low temperature region is shown in (b) in a larger
scale. To distinguish the curves some data points are marked by
symbols. The error bars are smaller than the symbols.\label{Fig1}}
\end{figure}

\begin{figure}
\caption[*]{Renormalization group improved field theory (solid lines)
[Eq. (\ref{1susc})] for different cut-off-temperatures $T_{0}$ versus
Bethe ansatz data \cite{eggert} and QMC results for $\chi(T)$ at low
temperature. The error bars are smaller than the
symbols.\label{Fig2}}
\end{figure}

\begin{figure}
\caption[*]{Susceptibility per rung of the 3-leg-ladder $\chi_{3}$
(dashed lines) for different ratios $J/J_{\perp}$ and of the single
chain $\chi_{1}$ in function of the temperature. The inset shows the
low temperature region in a larger scale. To distinguish the curves
some data points are marked by symbols. The error bars are smaller
than the symbols.\label{Fig3}}
\end{figure}

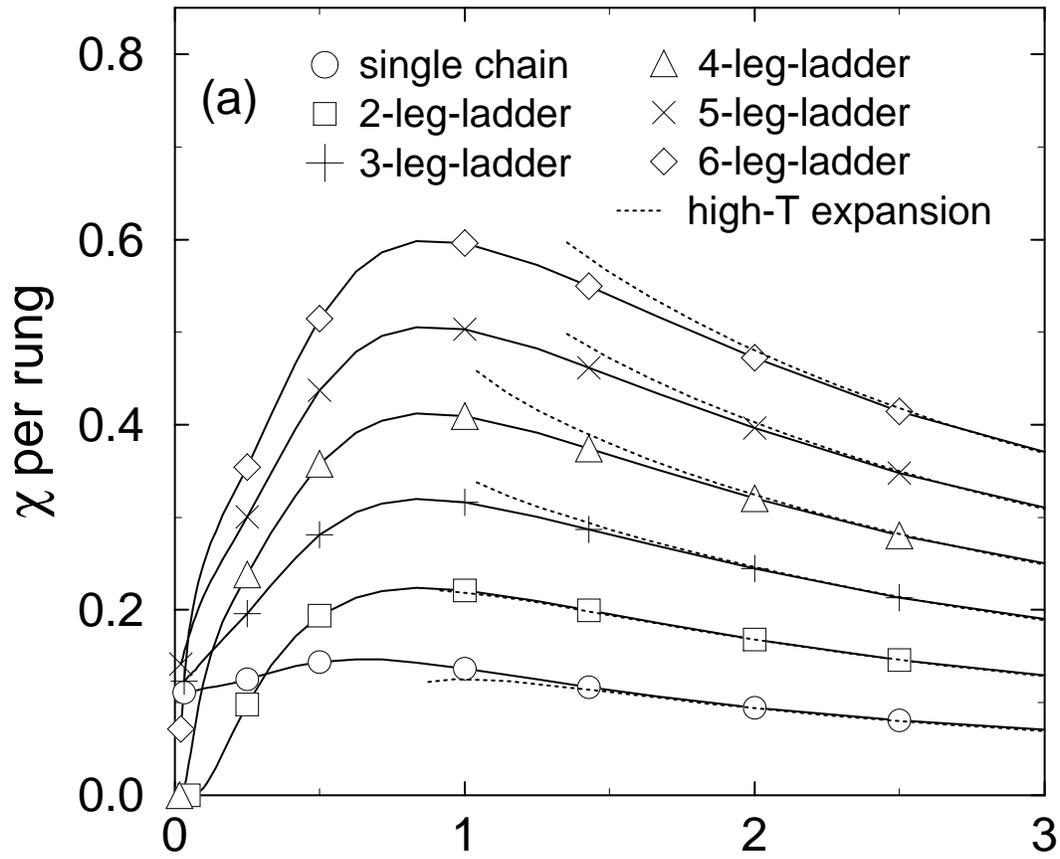
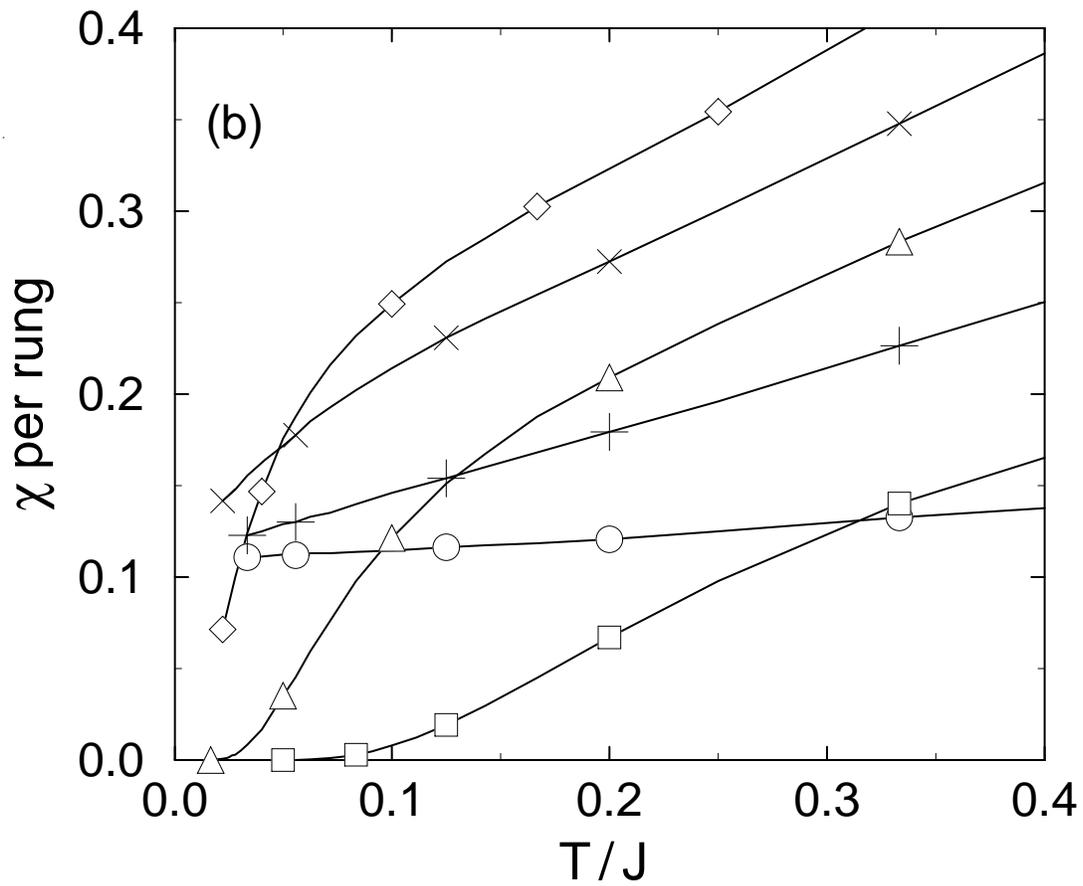
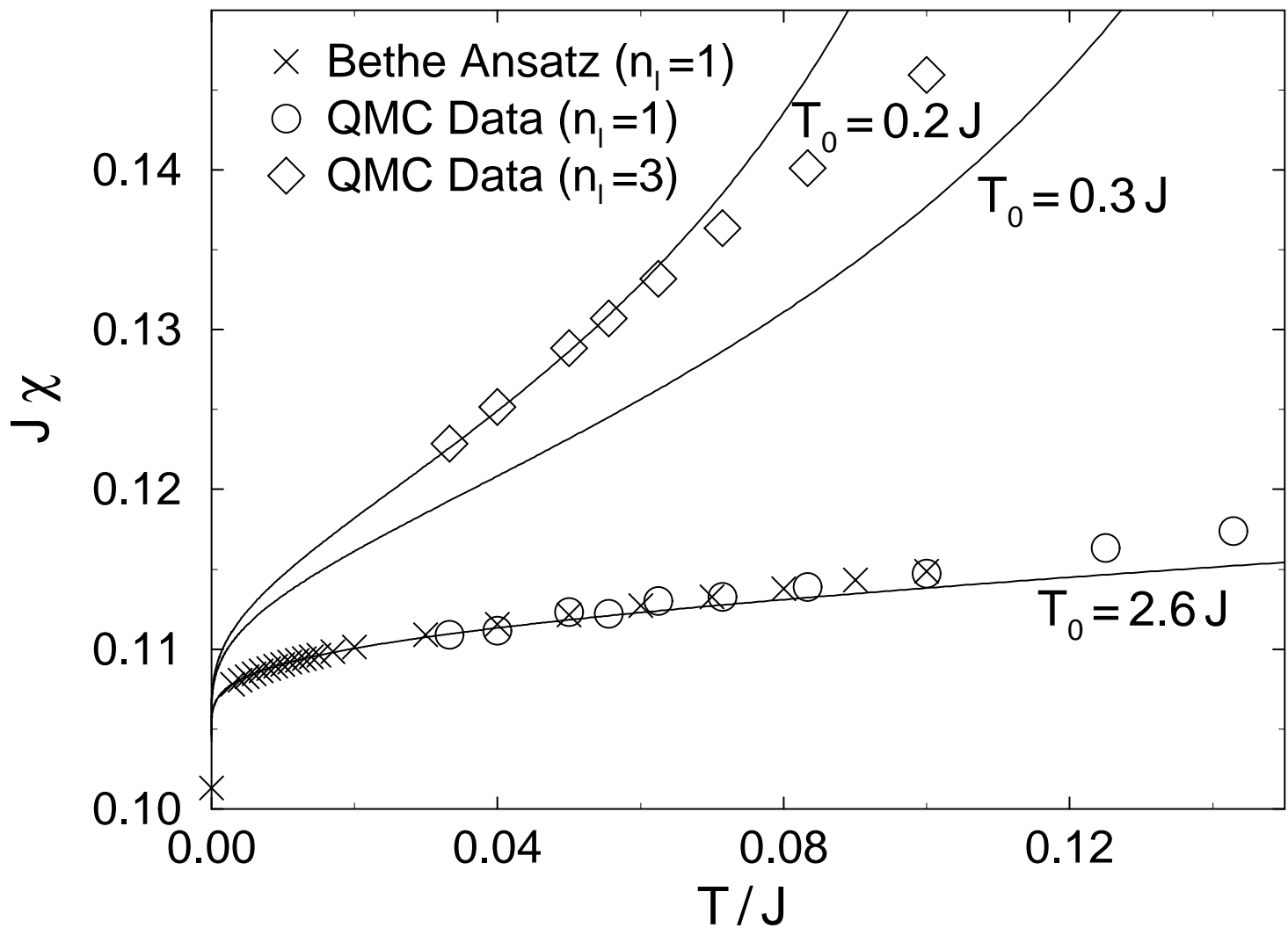
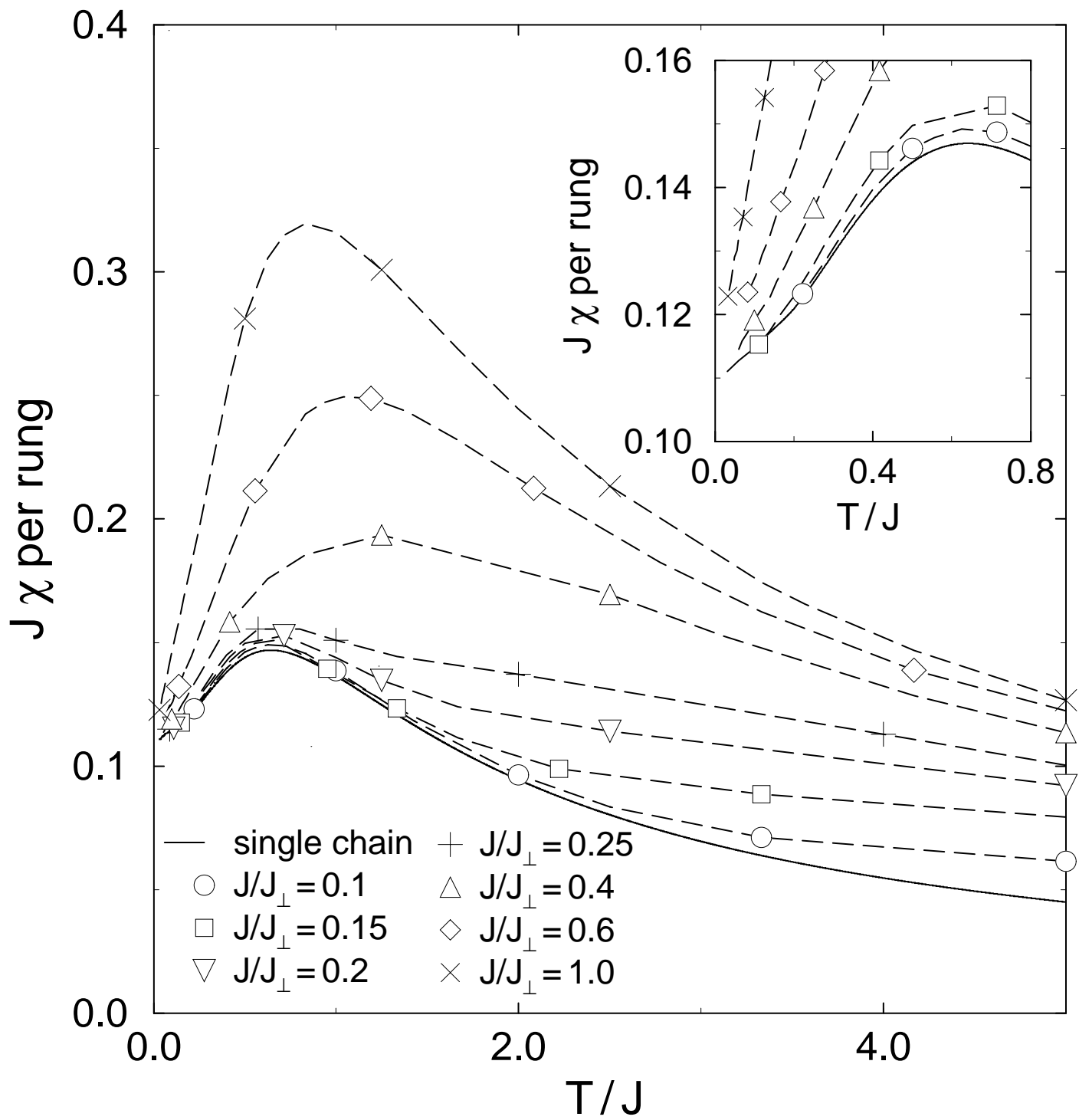
\end{document}